\documentclass[twocolumn, 5p]{elsarticle}

\usepackage{amsmath,amssymb}
\usepackage{subcaption}
\usepackage{graphicx}
\usepackage{hyperref}
\usepackage{multirow}
\usepackage{url}
\usepackage{color}
\usepackage{colortbl}
\usepackage{tabularx}
\usepackage{xspace}
\usepackage{booktabs}
\usepackage{listings}

\usepackage{algorithm}
\usepackage{algorithmicx}
\usepackage{amsmath,amssymb}
\usepackage{algpseudocode}

\usepackage{layouts}
\usepackage{wrapfig}

\usepackage{paralist}        

\usepackage{balance}          
\usepackage{multirow}         
\usepackage{multicol}     
\usepackage[font=small]{caption}         

\usepackage{comment}
\usepackage{siunitx}
\usepackage{hyperref}

\makeatletter
\def\ps@pprintTitle{%
	\let\@oddhead\@empty
	\let\@evenhead\@empty
	\def\@oddfoot{
		\textcopyright \ 2019. Licensed under the \href{https://creativecommons.org/licenses/by-nc-nd/2.0/}{CC-BY-NC-ND} license.\hfill
	}%
	\let\@evenfoot\@oddfoot}
\makeatother

\newcommand{\ie}{{i.e.}\xspace}
\newcommand{\eg}{{e.g.}\xspace}
\newcommand{\etal}{{et al.}\xspace}

\definecolor{dark-gray}{gray}{0.15}
\definecolor{black-gray}{gray}{0.2}
\definecolor{lightgray}{rgb}{.9,.9,.9}
\definecolor{darkgray}{rgb}{.4,.4,.4}
\definecolor{purple}{rgb}{0.8, 0.12, 0.52}
\definecolor{bluee}{rgb}{0.4, 0.4, 0.8}

\lstdefinelanguage{Dalvik}{
	alsoletter=?!-(),
	alsodigit=\$\%&*+./:;-><=>@^_~,
	keywords={invoke-virtual, invoke-static, invoke-direct, Landroid/app/admin/DevicePolicyManager;->, Ljavax/crypto/CipherOutputStream;->, Ljava/io/FileInputStream;->,
	Feature,List,Code,Vector},
	otherkeywords={lockNow,resetPassword,flush,close,read},	
	keywordstyle=\color{purple}\bfseries,
	ndkeywords={class, export, boolean, throw, implements, import, this,null},
	ndkeywordstyle=\color{darkgray}\bfseries,
	deletekeywords ={(Ljava/lang/}
	identifierstyle=\color{black},
	sensitive=false,
	comment=[l]{//},
	morecomment=[s]{/*}{*/},
	commentstyle=\color{purple}\ttfamily,
	stringstyle=\color{bluee}\ttfamily,
	morestring=[b]',
	morestring=[b]"
}

\lstdefinelanguage{Java}{
	alsoletter=?!-(),
	alsodigit=\$\%&*+./:<=>@^_~,
	keywords={public, throws, for, int, if, return, new},
	otherkeywords={this},	
	keywordstyle=\color{purple}\bfseries,
	ndkeywords={public,throws},
	ndkeywordstyle=\color{darkgray}\bfseries,
	deletekeywords ={(Ljava/lang/}
	identifierstyle=\color{black},
	sensitive=false,
	comment=[l]{//},
	morecomment=[s]{/*}{*/},
	commentstyle=\color{purple}\ttfamily,
	stringstyle=\color{bluee}\ttfamily,
	morestring=[b]',
	morestring=[b]"
}

\begin{document}

\begin{frontmatter}
	\title{On the Effectiveness of System API-Related Information\\for Android Ransomware Detection}
	
	\author[labelunica]{Michele Scalas\corref{correspondingauthor}}
	\address[labelunica]{Department of Electrical and Electronic Engineering, University of Cagliari,
		Piazza d’Armi 09123, Cagliari, Italy}
	\ead{michele.scalas@unica.it}
	\author[labelunica]{Davide~Maiorca\corref{coauthors}}%
	\ead{davide.maiorca@diee.unica.it}
	\author[labelpisa]{Francesco Mercaldo\corref{coauthors}}
	\address[labelpisa]{Institute of Informatics and Telematics, National Research of Council, Pisa, Italy}
	\ead{francesco.mercaldo@iit.cnr.it}
	\author[labelsannio]{Corrado Aaron Visaggio\corref{coauthors}}
	\address[labelsannio]{Department of Engineering, University of Sannio, Benevento, Italy}
	\ead{visaggio@unisannio.it}
	\author[labelpisa]{Fabio~Martinelli\corref{coauthors}}%
	\ead{fabio.martinelli@iit.cnr.it}
	\author[labelunica]{Giorgio Giacinto\corref{coauthors}}%
	\ead{giacinto@unica.it}
	
	\cortext[correspondingauthor]{Corresponding author}

	\begin{abstract}
		Ransomware constitutes a significant threat to the Android operating system. It can either lock or encrypt the target devices, and victims are forced to pay ransoms to restore their data. Hence, the prompt detection of such attacks has a priority in comparison to other malicious threats. Previous works on Android malware detection mainly focused on Machine Learning-oriented approaches that were tailored to identifying malware families, without a clear focus on ransomware. More specifically, such approaches resorted to complex information types such as permissions, user-implemented API calls, and native calls. However, this led to significant drawbacks concerning complexity, resilience against obfuscation, and explainability. To overcome these issues, in this paper, we propose and discuss learning-based detection strategies that rely on System API information. These techniques leverage the fact that ransomware attacks heavily resort to System API to perform their actions, and allow distinguishing between generic malware, ransomware and goodware.  
		We tested three different ways of employing System API information, \ie, through packages, classes, and methods, and we compared their performances to other, more complex state-of-the-art approaches. The attained results showed that systems based on System API could detect ransomware and generic malware with very good accuracy, comparable to systems that employed more complex information. Moreover, the proposed systems could accurately detect novel samples in the wild and showed resilience against static obfuscation attempts. Finally, to guarantee early on-device detection, we developed and released on the Android platform a complete ransomware and malware detector (R-PackDroid) that employed one of the methodologies proposed in this paper.    
	\end{abstract}
	
\end{frontmatter}

\section{Introduction}
\label{sec:intro}
The term \texttt{ransomware} refers to attacks that lock the victim's device or encrypt its data, by asking a sum of money to restore the compromised functionality. Despite the increasing diffusion of cloud-based technologies, users still store the majority of their data directly on their devices.  For this reason, such attacks are particularly devastating, as they could destroy sensitive data of private users and companies (which often neglect to make backups of sensitive data). According to Symantec, the number of ransomware variants increased in $2017$ by $46\%$, with massive outbreaks such as the one concerning Ukrainian companies (\texttt{Petya/NotPetya}). Hence, it is not surprising to see that the same trend applied to mobile ransomware, with more than $42,000$ samples blocked in $2017$~\cite{symantec-report18}.   

Mobile ransomware typically features different characteristics in comparison to its X86 counterpart. As performing data encryption typically requires high-level privileges (especially to write on areas that are directly managed by the kernel), most attacks only lock the target device by making victims believe that their data are encrypted, or by warning them that the police currently control them for their actions (a strategy directly inspired by \emph{scareware}-based approaches).

To counteract such attacks, Machine Learning has been increasingly used (especially combined with static analysis) both by researchers and anti-malware companies, either to perform direct detection or to generate signatures. While the goal of static detection systems is often to discriminate between generic malware and legitimate files (as in~\cite{demontis17-tdsc,rieck14-drebin,chen16-asiaccs,ahmadi17-cdmake}), some recently-released ones focus on further identifying malware families, in particular ransomware-related ones (also known as \emph{ransomware-oriented detection})~\cite{andronio2015heldroid,zheng16-securecomm,garcia18-tosem}. The reason for such a choice is that ransomware infections may lead to permanent data loss, making their early detection critical.

The main characteristic of systems to detect malware families is that they rely on different types of information extracted from multiple parts of the apps (\eg, bytecode, manifest, native libraries, and so forth~\cite{rieck14-drebin,andronio2015heldroid,zheng16-securecomm,chen16-asiaccs,garcia18-tosem}), which leads to use large amounts of features (even hundreds of thousands). While this approach is tempting and may seem to be effective against the majority of attacks in the wild, it features various limitations. First, it is unclear which features are essential (and needed) for classification, an aspect that worsens the overall explainability of the system (\ie, why the system makes mistakes and how to fix them). Second, increasing the types of features extends the degrees of freedom of a skilled attacker to perform targeted attacks against the learning algorithm. For example, it would be quite easy to mask a specific IP address, if the attacker understood that this has a vital role for detection~\cite{demontis17-tdsc}. Finally, the computational complexity of such systems is enormous, which makes them unfeasible to be practically used in mobile devices, an important aspect to guarantee offline, early detection of these attacks. 

In a previous work~\cite{maiorca17-sac}, we proposed a detection methodology (\texttt{R-PackDroid}) that allowed to discriminate between ransomware, generic malware, and legitimate files by focusing on a small-sized feature set, \ie, System API packages. The idea of our work was to overcome the limitations described above by showing that it was possible to solve a machine learning problem with a limited number of features of the same type. However, System API-based information does not only include packages but also classes and methods (particularly employed in other works, especially mixed with other feature types \cite{rieck14-drebin,garcia18-tosem}) that better define the behavior of APIs. Intuitively, using finer-grained information leads to better accuracy and robustness in comparison to other approaches. In this paper, we explore such a possibility by progressively refining the System API-based information employed in our previous work~\cite{maiorca17-sac}. In particular, \emph{we inspected the capabilities of multiple types of System API-related information to discriminate ransomware from malware and goodware}. More specifically, we aimed to provide an answer to the following Research Questions:    

\begin{itemize}
	\item \textbf{RQ 1.} Does the use of finer-grained information related to System API (\ie, classes and methods) improve detection performances in comparison to more general System API packages? 
	\item \textbf{RQ 2.} Is System API-based information suitable to detect novel attacks in the wild?
	\item \textbf{RQ 3.} Does using System API-based information provide comparable performances to other approaches that employ multiple feature types?
	\item \textbf{RQ 4.} Is System API-based information resilient against obfuscation attempts?
\end{itemize}

To answer such Research Questions, we explored three types of System API-based information: the first one only used information related to System API packages (as already shown in~\cite{maiorca17-sac}), the second one analyzed System API classes, and the third one employed information related to System API methods. We evaluated the performances of the three systems on a wide range of ransomware, malware and goodware samples in the wild (including previously unseen data). Moreover, we tested all systems against a dataset of ransomware samples that have been obfuscated with multiple techniques (including class encryption).

The attained results showed that all System API-based techniques provided excellent accuracy at detecting ransomware and generic malware in the wild, by also showing capabilities of predicting novel attacks and resilience against obfuscation. More specifically, using finer-grained information even improved the accuracy at detecting previously unseen samples, and provided more reliability against obfuscation attempts. From a methodological perspective, such results demonstrate that it is possible to develop accurate systems by strongly reducing the complexity of the examined information and by selecting feature types that represent how ransomware attacks behave. 

Finally, to demonstrate the practical suitability of System API-based approaches on Android devices, we ported to Android \texttt{R-PackDroid} (the package-based strategy originally proposed in~\cite{maiorca17-sac} and further explored in this work). Our application, which can detect both ransomware and generic malware in the wild, shows that methodologies based on System API can be implemented with good computational performances even in old phones, and its a demonstration of a full working prototype being deployed on real analysis environments. \texttt{R-Packdroid} can be downloaded for free from the Google Play Store\footnote{\url{http://pralab.diee.unica.it/en/RPackDroid}}.

With this work, we claim that it is possible to create effective, deployable, and reasonably secure approaches for ransomware and malware detection by only using specific feature types. Hence, we believe that the attention of research should be shifted to finding effective and explainable feature types to make detection even more accurate and robust.

\textbf{Paper structure.} Section~\ref{sec:background} provides the basic concepts of Android apps; Section~\ref{sec:ransomware} discusses the essential characteristics of Android ransomware and describes the key-intuitions behind using System API calls as critical information; Section~\ref{sec:relwork} provides a description of the related work in the field;  Section~\ref{sec:method} describes the employed detection methodologies; Section~\ref{sec:eval} illustrates the experimental results attained with all the methodologies, as well as a comparison between our systems and other approaches in the wild; Section~\ref{sec:implementation} describes the implementation details of \texttt{R-PackDroid} and its computational performances;  Section~\ref{sec:discussion} discusses the limitations of our work, which is finally concluded by Section~\ref{sec:conclusions}.

\section{Background on Android}
\label{sec:background}

Android applications are zipped \texttt{.apk} (i.e., Android application package) archives that contain the following elements: \emph{(i)} The \texttt{AndroidManifest.xml} file, which provides the application package name, and lists its basic components, along with the permissions that are required for specific operations; \emph{(ii)} One or more \texttt{classes.dex} files, which are the true executable of the application, and which contain all the implemented classes and methods (in the form of Dalvik bytecode) that are executed by the app. This file can be disassembled to a simplified format called \texttt{smali}; \emph{(iii)} Various \texttt{.xml} files that characterize the application layout; \emph{(iv)} External resources that include, among others, images and native libraries.

Although Android applications are typically written in \texttt{Java}, they are compiled to an intermediate byte-code format called \texttt{Dalvik} (which is further referred to as \texttt{DexCode}), whose instructions are contained in the \texttt{classes.dex} file. This file is then further parsed at install time and converted to native ARM code that is executed by the Android RunTime (ART). This technique allows to greatly speed up execution in comparison to the previous runtime (\texttt{dalvikvm}, available till Android $4.4$), in which applications were executed with a just-in-time approach (during installation, the \texttt{classes.dex} file was only slightly optimized, but not converted to native code).

\section{Android Ransomware}
\label{sec:ransomware}

The key point presented in this work is that the static extraction of System API-based information can be effective at detecting ransomware. More specifically, System APIs encapsulate many of the key actions performed by such attacks. To better reinforce this concept, in the following, we describe the basic actions performed by Android ransomware. 
The majority of ransomware-based attacks for Android are based on the goal of \emph{locking} the device screen while asking the victim for money in order to unlock it. According to the taxonomy proposed by \cite{chen18-tifs}, there are multiple ways to do so: \emph{(i)} by resorting to a \emph{hijacking} activity (\ie, a screen that the user visualizes and with which she can interact) that is continuously shown; \emph{(ii)} by setting up specific parameters of specific API calls; \emph{(iii)} by disabling certain buttons, such home or back. 

Locking is generally preferred to other data encryption strategies because it does not require to operate on high-privileged data. Indeed, accessing specific areas of the Android internal memory would only be possible with root permissions. Conversely, locking the device does not require particularly high privileges, and would allow the attacker to ensure his goal (\ie, scaring the victim) with minimum effort. The majority of locking screens show the victim writings and images related to police activities or pornographic material. There are, however, samples that also perform data encryption. According to \cite{chen18-tifs}, only four ransomware families possess the ability of encrypting data: \texttt{Simplocker}, \texttt{Koler}, \texttt{Cokri} and \texttt{Fobus}. In particular, some of these families employ a customized encryption algorithm, while others resort to standard algorithms. 

As locking and encryption actions require the use of multiple functions that involve core functionalities of the system (\eg, managing entire arrays of bytes, displaying activities, manipulating buttons and so on), \emph{attackers tend to use functions that directly belong to the Android System API}. It would be extremely time consuming and inefficient to build new APIs that perform the same actions as the original ones.

As an example of this behavior, consider the \texttt{DexCode} snippet provided by Listing~\ref{listing:locker}, belonging to a \emph{locker-type} ransomware\footnote{MD5: \texttt{0cdb7171bcd94ab5ef8b4d461afc446c}}. In this example, it is possible to observe that the two function calls (expressed by \texttt{invoke-virtual} instructions)  that are actually used to lock the screen (\texttt{lockNow}) and reset the password (\texttt{resetPassword}) are System API calls, belonging to the class \texttt{DevicePolicyManager} and to the package \texttt{android/app/admin}. The same behavior is provided by Listing \ref{listing:crypt}, which shows the encryption function employed by a \emph{crypto-type} ransomware sample\footnote{MD5: \texttt{59909615d2977e0be29b3ab8707c903a}}. Again, the functions to manipulate the bytes to encrypt belong to the System API (\texttt{read} and \texttt{close}, belonging to the \texttt{FileInputStream} class of the \texttt{java/io} package; \texttt{flush} and \texttt{close}, belonging to the \texttt{CipherOutputStream} class of the \texttt{javax/crypto} package).

In an Android application, based on Java, multiple methods are associated with classes that belong to packages. Because of these characteristics, it is possible to encode and represent System API information by either using packages, classes, or methods. More specifically, methods and classes better detail the functionality performed by the single API, but their number is significantly higher in comparison to packages. Hence, a solution that would employ the analysis of API methods would be far more complex than one that analyzes packages.

\begin{lstlisting}[language={Dalvik}, xleftmargin=0em, label=listing:locker, caption={Part of the onPasswordChanged() method belonging to a locker-type ransomware sample.},float=t]

invoke-virtual {v9}, Landroid/app/admin/DevicePolicyManager;->lockNow()V
move-object v9, v0
move-object v10, v1

...

move-result-object v9
move-object v10, v7
const/4 v11, 0x0
invoke-virtual {v9, v10, v11}, Landroid/app/admin/DevicePolicyManager;->resetPassword(Ljava/lang/String;I)Z

\end{lstlisting} 

\begin{lstlisting}[language={Dalvik}, xleftmargin=0em, label=listing:crypt, caption={Parts of the encrypt() method belonging to an encryption-type ransomware sample.},float=t]

Ljava/io/FileInputStream;->read([B)I
move-result v0
const/4 v5, -0x1
if-ne v0, v5, :cond_0
invoke-virtual {v1}, Ljavax/crypto/CipherOutputStream;->flush()V
invoke-virtual {v1}, Ljavax/crypto/CipherOutputStream;->close()V
invoke-virtual {v3}, Ljava/io/FileInputStream;->close()V

\end{lstlisting}

\section{Related Work}
\label{sec:relwork}

Most of Android malware detectors typically discriminate between malicious and benign apps, and we refer to them as \emph{generic malware-oriented} detectors. However, as the scope of this work is mostly oriented to ransomware detection, this Section will be mainly focused on describing systems that aim to detect such attacks (\emph{ransomware-oriented} detectors) specifically. A brief description of the other detectors will be provided at the end of this Section.

The most popular and publicly available \emph{ransomware-oriented} detector is \texttt{HelDroid}, proposed by Andronio \etal~\cite{andronio2015heldroid}. This tool includes a text classifier (based on NLP features) that works on suspicious strings used by the application, a lightweight \texttt{smali} emulation technique to detect locking strategies, and the application of taint tracking for detecting file-encrypting flows. The system has then been further expanded by Zheng \etal~\cite{zheng16-securecomm} with the new name of \texttt{GreatEatlon} and features significant speed improvements, a multiple-classifier system that combines the information extracted by text- and taint-analysis, and so forth. %
However, the system is still computationally demanding and it still strongly depends on a text classifier: the authors trained it on generic threatening phrases, similar to those that typically appear in ransomware or scareware.  This strategy can be easily thwarted by employing, \eg, string encryption~\cite{maiorca15-cose}. Moreover, it strongly depends on the presence of a language dictionary for that specific ransomware campaign. 

\normalsize
\begin{table*}[ht]
	\begin{center}
		\caption{An overview of the current state-of-the-art, ransomware-oriented approaches.}
		\label{sec:relwork:tab:works}
		\begin{tabular}
	{ |>{\bfseries}l  |c|c|c|c|c|c| }
	\hline
	\rowcolor[rgb]{.9,.9,1} \textbf{Work} & \textbf{Tool} & \textbf{Year} & \textbf{Static} & \textbf{Dynamic} & \textbf{Machine-Learning} & \textbf{Available}  \\
	\hline	
	Chen \etal~\cite{chen18-tifs} 				& RansomProber & $2018$ & & \checkmark & & \\ \hline
	Cimitille \etal~\cite{cimitile2017talos} 	& Talos 	   & $2017$ & \checkmark & & &\\ \hline
	Gharib \etal~\cite{Gharib2017} 				& Dna-Droid    & $2017$ & \checkmark & \checkmark &  \checkmark &\\ \hline
	Song \etal~\cite{song2016effective} 		& / 		   & $2016$ & & \checkmark & & \\ \hline
	Zheng \etal~\cite{zheng16-securecomm} 		& GreatEatlon  & $2016$ & \checkmark &  & \checkmark & \checkmark \\ \hline
	Yang \etal~\cite{yang2015automated} 		& / 		   & $2015$ & & \checkmark & & \\ \hline
	Andronio \etal~\cite{andronio2015heldroid}  & HelDroid 	   & $2015$ & \checkmark & & \checkmark & \checkmark \\ \hline   
\end{tabular}
	\end{center}
\end{table*}

Yang \etal~\cite{yang2015automated} proposed a tool to monitor the activity of ransomware by dumping the system messages log, including stack traces. Sadly, no implementation has been released for public usage.

Song \etal~\cite{song2016effective} proposed a method that aims to discriminate between ransomware and goodware using process monitoring. In particular, they considered system-related features representing the I/O rate, as well as the CPU and memory usage. The system has been evaluated with only one ransomware sample developed by the authors, and no implementation is publicly available. 

Cimitille \etal~\cite{cimitile2017talos} introduced an approach to detect ransomware that is based on formal methods (by using a tool called \texttt{Talos}), which help the analyst identify malicious sections in the app code. In particular, starting from the definition of payload behavior, the authors manually formulated logic rules that were later applied to detect ransomware. Unfortunately, such a procedure can become extremely time-consuming, as an expert should manually express such rules. 

Gharib \etal~\cite{Gharib2017} proposed \texttt{Dna-Droid}, a static and dynamic approach in which applications are first statically analyzed, and then dynamically inspected if the first part of the analysis returned a suspicious result. The system uses Deep Learning to provide a classification label.  The static part is based on textual and image classification, as well as on API calls and application permissions. The dynamic part relies on sequences of API calls that are compared to malicious sequences, which are related to malware families. This approach has the drawback that heavily obfuscated apps can escape the static filter, thus avoiding to be dynamically analyzed. Finally, Chen \etal~\cite{chen18-tifs} proposed \texttt{RansomProber}, a purely dynamic ransomware detector which employs a set of rules to monitor different aspects of the app execution, such as the presence of encryption or anomalous layout structures. The attained results report a very high accuracy, but the system has not been publicly released yet (to the best of our knowledge).    

Table \ref{sec:relwork:tab:works} shows a comparison between the state-of-the-art methods for specifically detecting or analyzing Android ransomware. It is possible to observe that there is a certain balance between static- and dynamic-based methods. Some of them also resort to Machine-Learning to perform classification. Notably, only \texttt{HelDroid} and \texttt{GreatEatlon} are currently publicly available. 

Concerning \emph{generic malware-oriented} detectors, Arp \etal~\cite{rieck14-drebin} proposed \texttt{Drebin}, a machine learning system that uses static analysis to discriminate between generic malware and trusted files. They extracted various features from both the Manifest file and the Android executable, including IP addresses, suspicious API calls, permissions, and so forth. Tam \etal~\cite{tam15-ndss} introduced a system to perform dynamic analysis and detection of Android malware by analyzing the system calls performed by the application. Avdieenko \etal~\cite{avdieenko15-icse} used taint analysis to detect anomalous flows of sensitive data, a technique that allowed to detect novel malware samples without previous knowledge. Yang \etal~\cite{yang15-icse} analyzed malicious apps by defining and extracting the context related to security-sensitive events. In particular, the authors defined a model of context based on two elements: activation conditions (\ie, what makes specific events occur) and guarding conditions (\ie, the environmental attributes of a specific event).

Aresu \etal~\cite{aresu15-malcon} clustered Android malware by using the network HTTP traffic generated by those applications. Such clusters can be used to generate signatures that allow discriminating between malware and legitimate applications. Canfora \etal~\cite{canfora2016hmm} experimentally evaluated two techniques for detecting Android malware: the first one is based on Hidden Markov Model (HMM), and the second one exploits Structural Entropy. The attained results showed that both techniques could be suitable for Android malware detection. 

Chen \etal~\cite{chen16-asiaccs} proposed \texttt{StormDroid}, a static and dynamic machine-learning based system that extracts information from API-calls, permissions and behavioral features. Ahmadi \etal~\cite{ahmadi17-cdmake} proposed \texttt{IntelliAV}, a \emph{generic malware-oriented} detector that is publicly available. Such a detector provides a level of dangerousness for each app but does not directly specify the family nor the type of attack. 

Garcia \etal~\cite{garcia18-tosem} proposed \texttt{RevealDroid}, a static system for detecting Android malware samples and classifying them in families. The system employs features extracted from reflective calls, native APIs, permissions, and many other characteristics of the file. The attained results showed that \texttt{RevealDroid} was able to attain very high accuracy, resilience to obfuscation. However, the number of extracted features can be extremely high and depends on the training data. 

Zhang \etal~\cite{zhang14-asiaccs,zhang16-aisec} leveraged machine learning-based solutions that employed dependency graphs extracted by observing network events. In particular, in~\cite{zhang14-asiaccs}, they proposed a traffic analysis method that employed scalable algorithms for the detection of malware activities on a host. Such a detection was performed by exploring request-level traffic and the semantic relationships among network events. The attained results showed high accuracy at detecting spyware, DNS bot, and data exfiltrating malware. In~\cite{zhang16-aisec}, they employed a learning-based solution that analyzed information extracted from the dynamic analysis of the network events generated by Android malware. In particular, they profiled the traffic generated by benign applications and modeled (through graphs called \emph{triggering relation graphs}) the triggering relationship of the generated network events (\ie, how such events are related to each other) to identify anomalous, malicious ones. The authors demonstrated that these graphs could be particularly useful to detect suspicious network requests. 

Finally, for the sake of completeness, we mention here other recent works that have analyzed the topic of ransomware detection on X86 (in particular, on Windows platforms) by employing dynamic analysis techniques to perform early detection of the attack. Using such techniques avoid possible damages to the operating system and its files.~\cite{continella16-acsac,kharaz16-usenix,kolodenker17-asiaccs,huang17-ccs}.

\section{Methodologies}
\label{sec:method}

We now describe the general structure of systems that employ System API information to identify ransomware, also known as \emph{ransomware-oriented} detectors. While the majority of learning-based detection systems combine various types of information to detect as many attacks as possible, \emph{ransomware-oriented} detectors tailor their detection on a smaller set of information (System API) that is typically employed in ransomware. However, as System APIs are also widely used in generic malware and legitimate files, this information type also allows detecting other attacks that differ to ransomware. In this way, it is possible to create a powerful, wide-spectrum detector that features a much lower complexity in comparison to other approaches.  
Typically, such systems take as input an Android application, analyze it and return three possible outputs: \textbf{ransomware}, \textbf{generic malware} or \textbf{trusted}. The analysis is performed in three steps: 

\begin{itemize}
	\item \textbf{Pre-Processing}. In this phase, the application is analyzed to extract its \texttt{DexCode}. The required information is extracted by only inspecting the executable code and does not perform any analysis on other elements, such as the application Manifest. Only specific lines of code, which will be described later in this Section, will be sent to the next module. 
	\item \textbf{Feature Extraction (System API)}. In this phase, the code lines received from the previous phase are further analyzed to extract the related \emph{System API} information (either packages, classes, or methods). The \emph{occurrence} of such pieces of information is then counted, thus producing a vector of numbers (\emph{feature vector}) that is sent to a classifier.
	\item \textbf{Classifier}. Classification is carried out through a \emph{supervised approach}, in which the system is trained with samples whose label (\ie, benign, generic malware or ransomware) is known. Such technique has been used in previous works with excellent results \cite{rieck14-drebin,demontis17-tdsc,garcia18-tosem}. In particular, our approaches employ Random Forest classifiers, which are especially useful to handle multi-class problems, and which are widely used for malware detection. The complexity of such classifiers depends on the number of trees that compose them. Such a number must be optimized during the training phase. 
\end{itemize}

\begin{figure*}[htp]
	\centering
	\includegraphics[width=\textwidth]{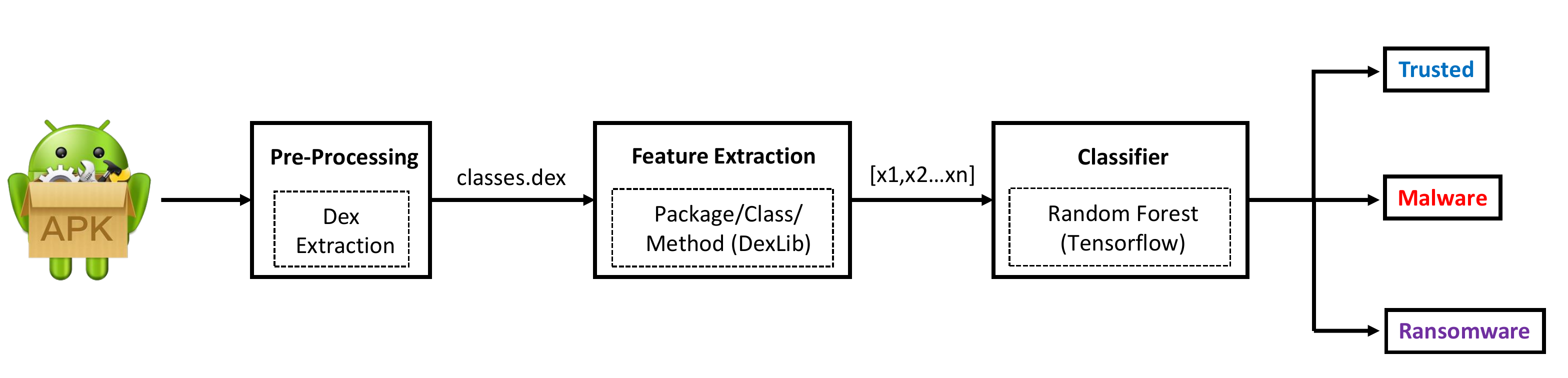}
	\caption{General Structure of a System API-based, ransomware-oriented system.}
	\label{sec:method:fig:pdfscheme}
\end{figure*}

The structure above is graphically represented in Figure \ref{sec:method:fig:pdfscheme}. In the following, we provide more details about each phase of the analysis, by focusing in particular on the type of features that can be extracted from the application.

\subsection{Preprocessing and Feature Extraction}
\label{sec:method:subsec:feat}

The general idea of the first two phases is performing static analysis of the Dalvik bytecode contained in the \textbf{classes.dex} file. The goal is retrieving the \emph{System API information} employed by the executable code of the application. The choice of System API information is related to two basic ideas:

\begin{itemize}
\item \textbf{Coherence with actions}. Most ransomware writers resort to System APIs to carry out memory- or kernel-related actions (for example, file encryption or memory management). Focusing on user-implemented APIs (as it happens, for example, with \texttt{Drebin} \cite{rieck14-drebin}) exposes the system to a risk of being evaded by simply employing different packages to perform actions. 
\item \textbf{Independence from Training}. System API calls are features independent of the training data that are used. As a consequence, it is less likely that applications are not correctly analyzed only because they employ never-seen-before APIs. 
\item \textbf{Resilience against obfuscation}. Using heavy obfuscation routines typically lead to injecting system API-based code in the executable, which can be extracted and analyzed, allowing to detect suspicious files. 
\end{itemize}

Pre-processing is hence easily performed by directly extracting the \textbf{classes.dex} file from the \texttt{.apk} app. Since \texttt{.apk} files are essentially zipped archives, such an operation is rather straight-forward.  

Once pre-processing is complete, the \texttt{classes.dex} file is further analyzed by the feature extraction module, which inspects the executable file for all \texttt{invoke}-type instructions (\ie, all instructions related to invocations) contained in the \texttt{classes.dex} code. Then, each invocation is inspected to extract the relevant API information for each methodology, according to a System API reference list that depends on the operating system version (in our case, Android Nougat - API level $25$ - a widely-used API set). Only the API elements that belong to the reference list are analyzed. In this paper, we consider three different methodologies, based on, respectively, package, class, and method extraction. If a specific API element is found, its occurrence value is increased by one. 

In the following, we provide a more detailed description of the methodologies employed in this paper, by referring to the example reported in Listing \ref{listing:as_example}. The code is parsed in three ways, according to each feature extraction strategy. For each example, we used a very small subset of the employed reference API.

\begin{itemize}
	
	\item \textbf{Packages Extraction.} In this methodology, we extract the occurrences of the System API packages (a total of $270$ reference features), in the same way of our previous work \cite{maiorca17-sac}. In the example of Listing \ref{listing:as_example}, we used a subset composed of three reference API packages: \texttt{java/io}, \texttt{java/crypto} and \texttt{java/lang}. The four \texttt{invoke} instructions are related to the \texttt{javax/crypto} and \texttt{java/io} packages, which are counted respectively twice. The \texttt{java/lang} package is never used in this snippet. Hence, its value is zero. 
	\item \textbf{Classes Extraction.} In this methodology, we extract the occurrences of the System API classes (a total of $4609$ reference features). Notably, such classes belong to the System API packages of the previous methodology (and, for this reason, their number is significantly higher than packages).  In the example of Listing \ref{listing:as_example}, we used a subset composed of two reference API classes: \texttt{java/io/FileInputStream} and  \texttt{javax/crypto/CypherOutputStream}, each of them appearing twice.
	\item \textbf{Methods Extraction.} In this methodology, we extract the occurrences of the System API methods (a total of $36148$ reference features). These methods belong to the System API classes of the previous methodology, leading to a very consistent number of features. This strategy is very similar to other ones employed by other systems (\eg. \cite{rieck14-drebin,garcia18-tosem}), which have used these features together with user-implemented APIs and other features. In the example of Listing \ref{listing:as_example}, we used a subset composed of four reference API methods: \texttt{java/io/FileInputStream/read}, \texttt{javax/crypto/CypherOutputStream/flush}, \texttt{javax/crypto/CypherOutputStream/close} and \texttt{java/io/FileInputStream/close}. Each API call appears only once. Note that, although there are two methods named \emph{close}, they belong to two different classes, and they are therefore considered as different methods.

\end{itemize}

\begin{lstlisting}[language={Dalvik}, xleftmargin=0em, label=listing:as_example, caption={An example of feature extraction by considering a small number of reference features.},float=t]

Code

Ljava/io/FileInputStream;->read([B)I
move-result v0
const/4 v5, -0x1
if-ne v0, v5, :cond_0
invoke-virtual {v1}, Ljavax/crypto/CipherOutputStream;->flush()V
invoke-virtual {v1}, Ljavax/crypto/CipherOutputStream;->close()V
invoke-virtual {v3}, Ljava/io/FileInputStream;->close()V

Feat. Vectors

Packages - [2 2 0]

Classes - [2 2]

Methods - [1 1 1 1]


\end{lstlisting}

\section{Experimental Evaluation}
\label{sec:eval}

In this Section, we report the experimental results attained from the evaluation of the three API-based strategies. Note that, for the sake of simplicity and speed, we did not run the experiments on Android phones, but on an X86 machine. However, we built a full, working implementation of one of the three approaches, which can be downloaded from the Google Play Store (see next Section). 

The rest of this Section is organized as follows: we start by providing an overview of the dataset employed in our experiments. Then, we describe the results attained by four evaluations. The first one aimed to establish the general performances of API-based approaches by considering random distributions of training and test samples. The second one aimed to show how API-based approaches behaved when analyzing samples released after the training data. The third one aimed to show a comparison between our API-based approaches and other systems that employed mixed features. Finally, we evaluated the resilience of API-based approaches against obfuscation techniques and evasion attacks.  

\subsection{Dataset}

\label{sec:eval:subsec:dataset}
In the following, we describe the dataset employed in our experiments. Without considering obfuscated applications (which are going to be discussed in Section \ref{sec:eval:subsec:exp2}), we obtained and analyzed \num{39157} apps, which are organized in the three categories we mentioned in Section \ref{sec:method}. 

\subsubsection{Ransomware}
\label{sec:eval:subsec:dataset:subsubsec:ransom}

The 3017 samples used for our ransomware dataset were retrieved from the \texttt{VirusTotal} service\footnote{\url{http://www.virustotal.com}} (which aggregates the detection of multiple anti-malware solutions) and from the \texttt{HelDroid} dataset\footnote{\url{https://github.com/necst/heldroid}} \cite{andronio2015heldroid}.
With respect to the samples obtained from \texttt{VirusTotal}, we used the following procedure to obtain the samples: \emph{(i)} we searched and downloaded the Android samples whose anti-malware label included the word \emph{ransom}; \emph{(ii)} for each downloaded sample, we extracted its family by using the \texttt{AVClass} tool~\cite{sebastian16-raid}, which essentially combines the various labels provided by anti-malware solutions to create a unique label that identifies the sample itself; \emph{(iii)} we considered only those samples whose family was coherent to ransomware behaviors, or was known to belong to ransomware. 

\normalsize
\begin{table}[!h]
	\begin{center}
		\caption{Ransomware families included in the employed dataset.}
		\label{sec:eval:subsec:dataset:subsubsec:exp1:tab:results}
			\begin{tabular}
				{ |>{\bfseries}l  |c| }
				\hline
				\rowcolor[rgb]{.9,.9,1} \textbf{Family} &  \textbf{Samples} \\
				\hline	
				Locker &  $752$\\ \hline
				Koler & $601$\\ \hline
				Svpeng & $364$\\ \hline
				SLocker & $281$\\ \hline
				Simplocker & $201$\\ \hline
				LockScreen & $122$\\ \hline
				Fusob & $120$\\ \hline
				Lockerpin & $120$\\ \hline
				Congur & $90$\\ \hline
				Jisut & $86$\\ \hline
				Other & $280$ \\ \hline
			\end{tabular}
	\end{center}
\end{table}

In general, our goal was obtaining a representative corpus of ransomware to ascertain the prediction capabilities of API-based techniques. For this reason, the dataset includes families that perform both device locking (such as \texttt{Svpeng} and \texttt{LockScreen}) and encryption (such as \texttt{Koler} and \texttt{SLocker}).  For a better description of the families above, please see Section \ref{sec:ransomware}. 

\subsubsection{Malware and Trusted}
\label{sec:eval:subsec:dataset:subsubsec:maltrusted}

We considered a dataset composed of \num{17744} Android malware samples that do not belong to the ransomware category, taken from the following sources: \emph{(i)} \texttt{Drebin} dataset, one of the most recent, publicly available datasets of malicious Android applications\footnote{\url{https://www.sec.cs.tu-bs.de/~danarp/drebin/}} (which also contains the samples from the Genome dataset ~\cite{zhou2012dissecting}); \emph{(ii)} \texttt{Contagio}, a popular free source of malware for X86 and mobile; \emph{(iii)} \texttt{VirusTotal}. These samples were chosen to verify whether even non-ransomware attacks could be detected with features that are particularly effective at classifying ransomware samples.  

In order to download trusted applications, we resorted to two data sources: \emph{(i)} we crawled the Google Play market using an open-source crawler\footnote{\url{https://github.com/liato/android-market-API-py}}; \emph{(ii)} we extracted a number of applications from the \texttt{AndroZoo} dataset \cite{allix16-msr}, which features a snapshot of the \texttt{Google Play} store, allowing to access applications without crawling the Google services easily. We obtained \num{18396} applications that belong to all the different categories available on the market. We chose to focus on the most popular apps to increase the probability of downloading malware-free apps.

\begin{figure*}[t]
	\centering
	\label{sec:eval:subsec:exp1:fig:roc_ransomware}
	\includegraphics[width=0.5\textwidth]{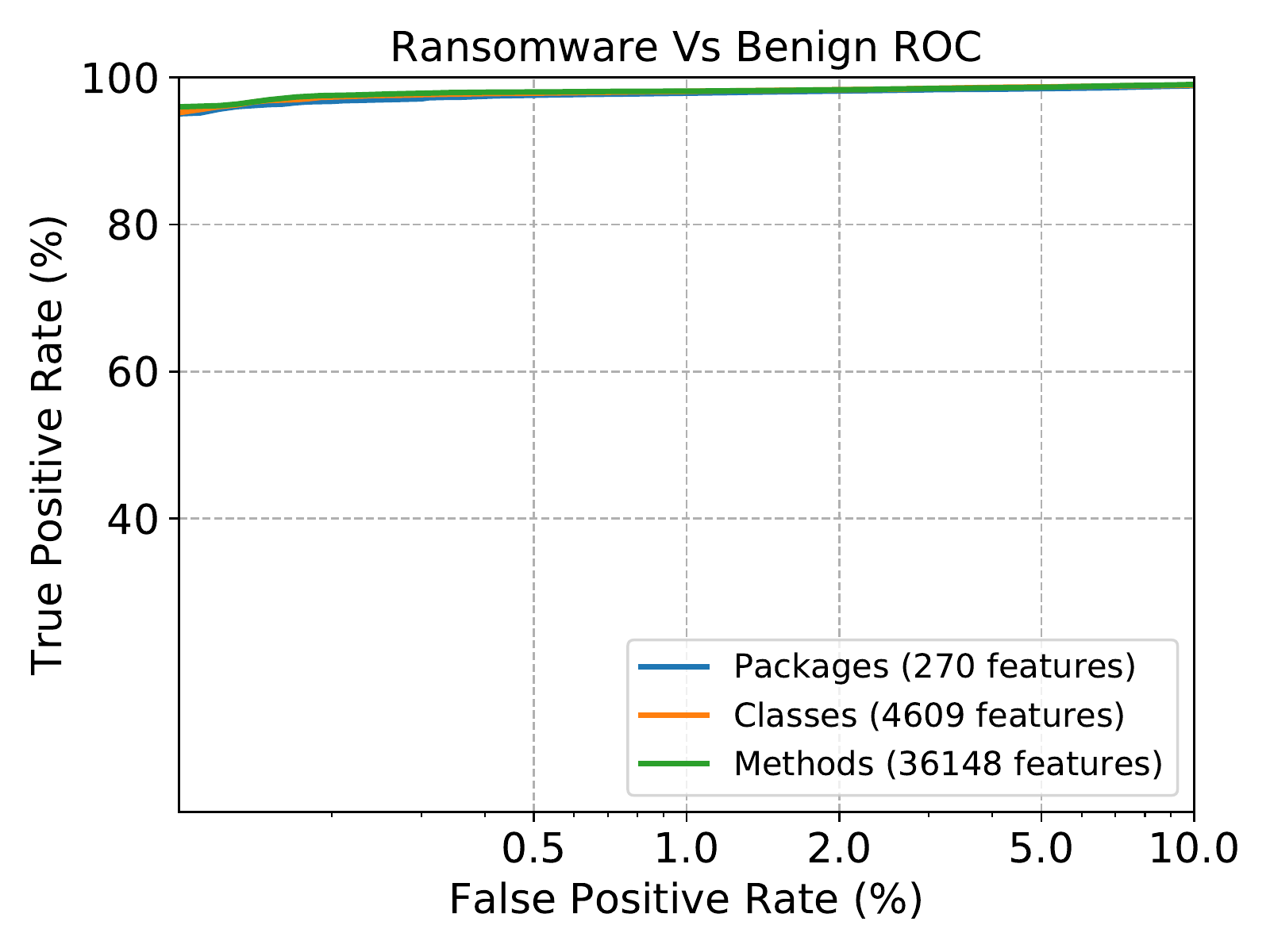}
	\label{sec:eval:subsec:exp1:fig:roc_malware}
	\includegraphics[width=0.49\textwidth]{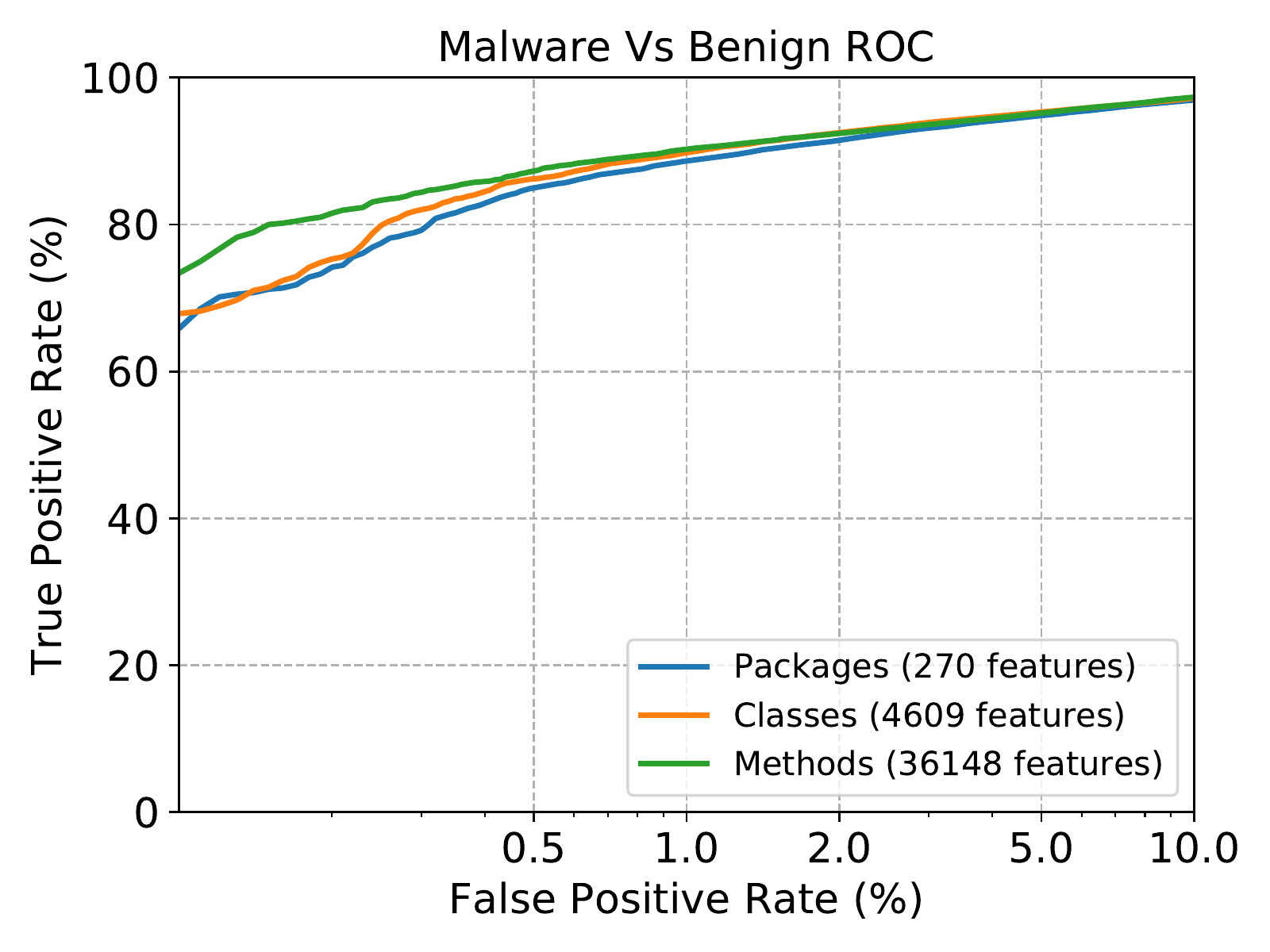}
	\begin{minipage}{0.45\textwidth}
		\subcaption{}
	\end{minipage}
	\begin{minipage}{0.45\textwidth}
		\subcaption{}
	\end{minipage}
	{\fontsize{8.5pt}{\baselineskip}\fontfamily{DejaVuSans-TLF}\selectfont
	\begin{minipage}{0.33\textwidth}
		\text{Top discriminant features (Packages)}
	\end{minipage}
	\begin{minipage}{0.33\textwidth}
		\text{Top discriminant features (Classes)}
	\end{minipage}
	\begin{minipage}{0.3\textwidth}
		\text{Top discriminant features (Methods)}
	\end{minipage}
	}
	\label{sec:eval:subsec:exp1:fig:feats}
	\includegraphics[width=\textwidth, trim={0 0 0 0.4cm}, clip]{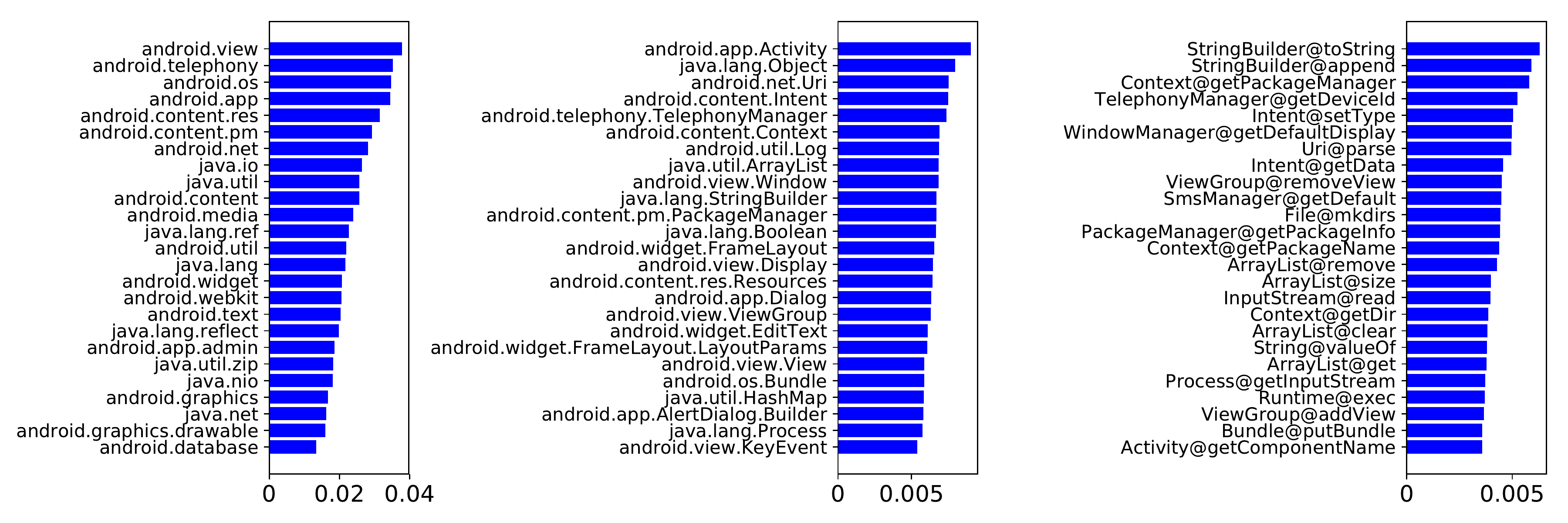}
	\begin{minipage}{0.33\textwidth}
		\subcaption{}
	\end{minipage}
	\begin{minipage}{0.33\textwidth}
		\subcaption{}
	\end{minipage}
	\begin{minipage}{0.3\textwidth}
		\subcaption{}
	\end{minipage}
	\caption{Results for the experiment $1$. In the first row, we report the ROC curves (averaged on $5$ splits) attained by the three System API methods for ransomware (a) and generic malware (b) against benign samples. In the second row, we report the top-$25$ features, ordered by the classifier information gain (calculated by averaging, for each feature, the gains that were obtained by training the $5$ splits), for each methodology ((c) for packages, (d) for classes, (e) for methods).}
	\label{sec:eval:subsec:exp1:exp1_results}
\end{figure*}

\subsection{Experiment 1: General Performances}
\label{sec:eval:subsec:exp1}

In this experiment, we evaluated the general performances of System API-Based methods (described in Section \ref{sec:method}) at detecting ransomware and generic malware. To do so, for each strategy, we randomly split our dataset by $50\%$, thus using the first half to train the system and the second half to evaluate the system. The number of trees of the random forests was evaluated by performing a $10$-fold cross-validation on the training data. We repeated the whole process $5$ times, and we averaged the results by also determining the standard deviation, in order to understand the dependence of the system on the training data. 

Considering the multi-class nature of the problem, we represented the results by calculating the ROC curve for each API-based strategy in two different cases:

\begin{itemize}
	\item \textbf{Ransomware against benign samples}. The crucial goal of our work is detecting ransomware attacks and, more importantly, \emph{to avoid them being considered as benign files}. A critical mistake would most likely compromise the whole device by locking it or encrypting its data. For this reason, it is essential to verify whether ransomware attacks can be confused with benign samples.
	\item \textbf{Generic malware against benign samples}. Even if System API-based strategies were employed to detect ransomware, they could also be used to classify generic malware (see Section \ref{sec:method}). Hence, the goal here is to verify, from a practical perspective, if System API-based information can correctly detect other non-ransomware attacks and distinguish them from legitimate files.
\end{itemize}

Results are reported in Figure \ref{sec:eval:subsec:exp1:exp1_results}. Parts (a) and (b) show the ROC curves that describe the performances attained on ransomware and generic malware detection by the three System API-based methods (packages, classes, methods). By observing these curves, we can deduce the following facts:

\begin{enumerate}
	\item All System API-based techniques were able to precisely detect more than $97\%$ of ransomware samples with only $1\%$ of false positives. Because our dataset included a consistent variety of families, we claim that all strategies can detect the majority of ransomware families in the wild. Worth noting, there are no differences in results between using packages, classes or methods. This result means that, concerning general performances, using finer-grained features does not improve detection.
	\item All System API methods featured good accuracy with relatively low false positives (around $90\%$ at $1\%$, more than $95\%$ at $2\%$) at detecting generic malware. While using class-related features did not bring significant improvements to detection, using methods allowed for a $10\%$ improvement for false positive values inferior to $0.5\%$.  
\end{enumerate}

To better understand the results attained by our strategies, parts (c), (d) and (e) report a ranking of features used by the classifier for each strategy (respectively, packages, methods, and classes), according to their discriminant power. 
 The feature ranking is calculated according to the features Information Gain $IG$, given by the following formulation:
 
\begin{equation}
IG(T,a) = H(T) - H(T|a)
\end{equation}

where $H(T)$ is the overall entropy for the whole dataset $T$ and $H(T|a)$ is the average entropy obtained by splitting the set $T$ using the attribute $a$. The higher is the gain, the more relevant the feature is.

As a result, note how the information gain for each feature is not so high, meaning that the system does not particularly overfit on specific information and that the final decision is taken by considering a combination of multiple features. At the same time, each feature value is reduced, in comparison to packages, by one magnitude for classes and methods. In other words, using more features allows for distributing the importance of the analyzed information through more elements. This characteristic is two-faced: while it makes the overall behavior of the system less interpretable, it may increase the effort that an attacker has to make to evade the system. 

Analyzing the most discriminant methods can give a clearer idea of which information is used to classify applications. Features are related to string building (\eg, the \texttt{ToString} method), Array management (\eg, \texttt{ArrayList@size}, \texttt{ArrayList@remove}), creation of folders (\eg, \texttt{File@mkdirs}), SMS, URI and Layout management, and so forth. These features may be easily associated both to ransomware and malware behavior, and the same behavior is shown on classes and packages. 

A careful examination of the feature ranking may also help to understand why specific samples are regarded as false positives or false negatives. In particular, detection is performed by weighting the information provided by a combination of the most discriminant features. For this reason, in some cases, specific ransomware (or generic malware) samples may contain discriminant features that are distributed differently to the malicious training distribution. For example, the \texttt{toString} call may appear, on average, $5$ times on ransomware, but one specific sample may feature it only $1$ time. This phenomenon may occur for various reasons, including the possibility that an attacker may be using customized variants of System API information to avoid detection (see Section~\ref{sec:discussion}). Likewise, the techniques used to create the sample (\eg, repackaging) may have an impact on the distribution of the features. Further refinement of the feature list (or a change of the weights of specific feature) may help to reduce the amount of misclassified samples.

\subsection{Experiment 2: Temporal Performances}
\label{sec:eval:subsec:exp2}

In this experiment, we assessed the capabilities of System API-based methods at detecting ransomware samples that were first seen (according to the creation date of the \texttt{classes.dex} executable belonging to each application) \emph{after} the data that were used as training set. This assessment is useful to understand if, without constant upgrades to its training set, such methods would be able to detect novel, unseen ransomware samples. 

\begin{figure}[htp]
	\centering
	\includegraphics[width=\linewidth]{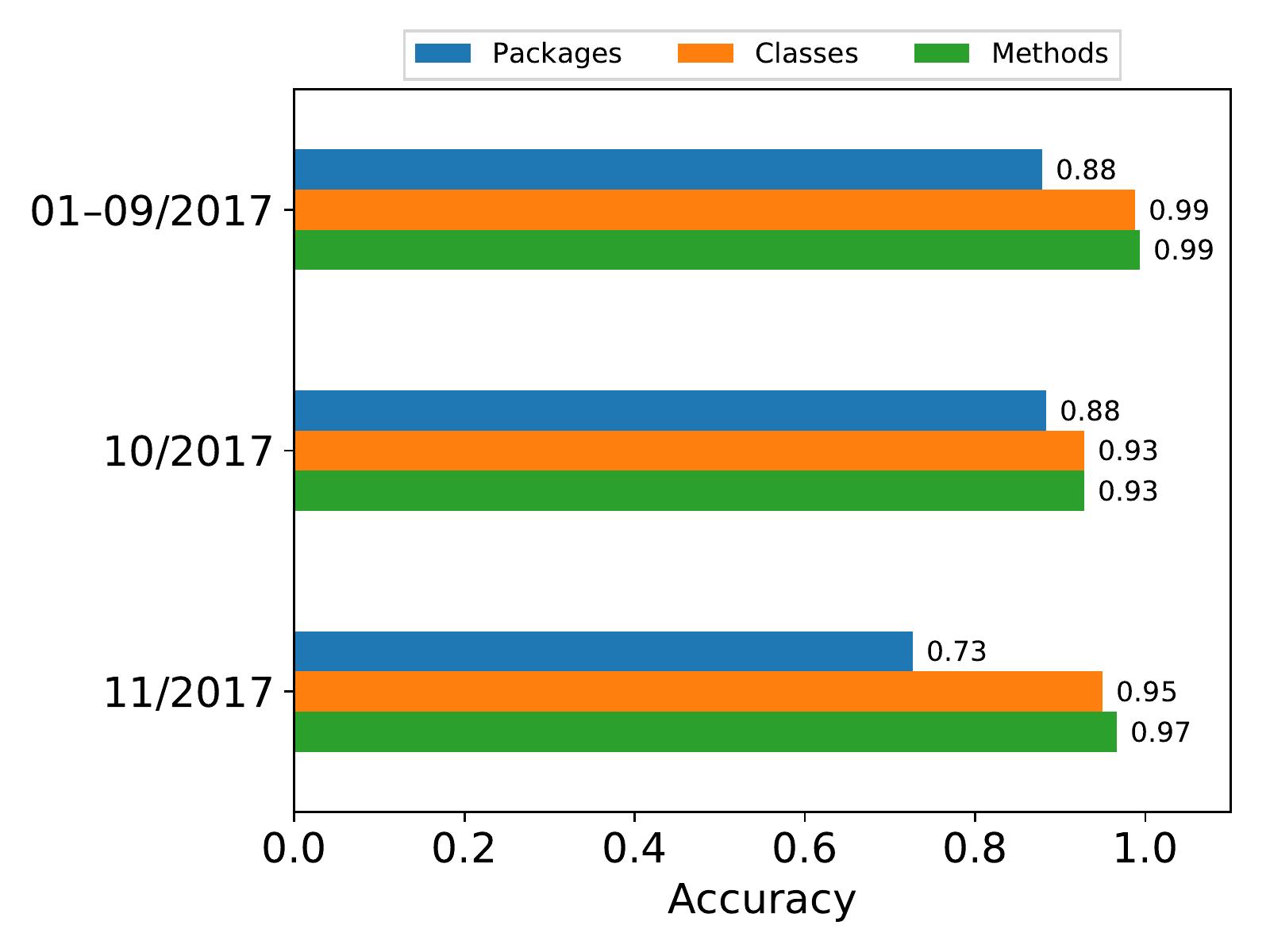}
	\caption{Results of the temporal evaluation for the System API-based strategies. The accuracy values are reported for the three System API-based detection. The training data belong to $2016$, while the test data is composed of ransomware released in different months of $2017$.}
	\label{sec:eval:subsec:exp2:fig:temporal_accuracy}
	
\end{figure} 

For this assessment, we included in the training set (along with \emph{all generic malware and trusted samples}) ransomware samples that were first seen before a date $D_{tr}$, and we tested our system on a number of ransomware samples that were released on a date $D_{te}$ for which $D_{te} > D_{tr}$ (we chose the ROC operating point of the system corresponding to a false positives value of $1$\%). We performed our tests by choosing different values of $D_{te}$, where $D_{tr}$ is December $31$st, $2016$. Concerning test data, we point out that the samples (which were extracted by the \texttt{VirusTotal} service) were unevenly distributed through the months. More specifically, the number of ransomware samples that were submitted to the \texttt{VirusTotal} service was significantly different for each month of $2017$. In particular, we could retrieve only a little amount of samples whose first release date was between January and September $2017$. Conversely, we could obtain a consistent amount of samples whose $D_{te}$ was October and November $2017$. Therefore, we grouped the samples gathered in subsequent months to obtain temporal ranges with similar amounts of testing data. We considered three main ranges for $D_{te}$: \emph{(i)} January to September $2017$; \emph{(ii)} October $2017$; \emph{(iii)} November $2017$.

Results are provided in Figure \ref{sec:eval:subsec:exp2:fig:temporal_accuracy}, which shows that by training the system with data retrieved in $2016$, class- and method-based strategies could accurately detect ransomware test samples released in $2017$. However, the package-based strategy struggled at detecting the test-set from November $2017$. Notably, in comparison with class- and method-based strategies, the package-based approach showed almost a $10$\% accuracy loss when analyzing samples released till October, and more than $20$\% accuracy loss for samples released in November. Conversely, the other two methods exhibited stable accuracy on each temporal range. This result is particularly interesting, as it shows that the prediction of novel samples can be significantly improved by employing finer-grained features. However, the results attained by package-based features are nevertheless encouraging, as they showed that even a reduced number of features could attain good performances at detecting novel attacks. Overall, this experiment further confirmed that System API-based strategies could predict new ransomware attacks with good accuracy, even on test data released. In this case, using finer-grained features brings a consistent advantage to detection. 

\subsection{Experiment 3: Comparison with Other Approaches}
\label{sec:eval:subsec:exp3}
This section proposes a comparison between System API-based strategies and other state-of-the-art approaches. We were particularly interested in comparing our approach to other publicly available ones, with a special focus on those who were specifically designed to detect ransomware. Additionally, we considered those approaches that, albeit not explicitly designed to detect ransomware, could tell if the analyzed sample is malicious or not. In particular, almost all of the analyzed tools (except for \texttt{Talos}~\cite{cimitile2017talos} - which does not employ Machine Learning) discriminate between two classes (malware and benign or ransomware and benign), while our approaches discriminate between three classes (ransomware, malware, and benign). Hence, it is interesting to observe how increasing the number of classes may impact the precision of the analysis. 

We performed a temporal comparison of all systems on the ransomware samples released in $2017$ (for a total of $512$ samples) by using as training (when possible) all data released until $2016$.

The state-of-the-art approach that is closest to what we proposed in this paper (while being publicly available\footnote{\url{https://github.com/necst/heldroid}}) is \texttt{GreatEatlon}~\cite{zheng16-securecomm}. Notably, it was not possible for us to control the trained model of the system (it was only possible to choose among a restricted set of classifiers), or to train it with new data. Nevertheless, the system was released in $2016$, meaning that data that was first seen in $2017$ was for sure not included in its training set. Although not specifically tailored to ransomware detection, we also tested the performances of \texttt{RevealDroid} (which is publicly available\footnote{\url{https://seal.ics.uci.edu/projects/revealdroid/}} ~\cite{garcia18-tosem}) on the same test data. In this case, we could train the system with the same data used in our systems, which allowed us to provide a fairer comparison. Finally, we also tested the performances of the Android version of \texttt{IntelliAV} (available on the Google Play Store) \cite{ahmadi17-cdmake}. As in \texttt{GreatEatlon}, we could not control the training data of the system. Moreover, as \texttt{IntelliAV} reports three levels of risk for each app (safe, suspicious, risky), we considered as malicious also the files that were labeled as suspicious by the system. 

As classifier for \texttt{GreatEatlon} we chose Stochastic Gradient Descent (SGD), since this was the classifier that best performed on our test samples. 
Concerning \texttt{RevealDroid}, we chose the linear SVM classifier, as this was the one that provided the best results in the original work ~\cite{garcia18-tosem}. \texttt{IntelliAV} only employed Random Forests. Results are reported in Table~\ref{sec:exp3:subsec:comparison:tab:results}.

\normalsize
\begin{table*}[htp]
	\begin{center}
		\caption{Detection performances for System API-based strategies, \texttt{GreatEatlon}, \texttt{IntelliAV}, \texttt{RevealDroid} and \texttt{Talos} on $512$ ransomware test files released in $2017$, by using training data from $2016$. We use the ND (Not Defined) to indicate that a certain tool cannot provide the corresponding label for the analyzed samples.}
		\label{sec:exp3:subsec:comparison:tab:results}
		\begin{tabular}
			{ |>{\bfseries}l  |c|c|c|}
			\hline
			\rowcolor[rgb]{.9,.9,1} \textbf{System} &  \textbf{Benign} & \textbf{Generic Malware} & \textbf{Ransomware}\\
			\hline
			Talos                 & $3$    & $0$   &
			$509$ \\ \hline
			System API (Methods)  & $7$    & $12$  & $493$ \\ \hline
			System API (Classes)  & $10$   & $15$  & $487$ \\ \hline
			System API (Packages) & $11$   & $32$  & $469$ \\ \hline
			GreatEatlon 		  & $118$  & $ND$   & $394$\\ \hline
			RevealDroid 		  & $0$    & $512$ & $ND$\\ \hline
			IntelliAV   		  & $18$   & $494$ & $ND$\\ \hline
		\end{tabular}
	\end{center}
\end{table*}

The attained results show that System API-based techniques obtained very similar performances to \texttt{RevealDroid} (which could only, however, classify samples either as malware or benign). Such results are particularly interesting if we consider that \texttt{RevealDroid} extracted a huge number of features (more than $700,000$) from multiple characteristics of the file, including native calls, permissions, executable code analysis, which also depended on the training data. With a much simpler set of information, we were able to obtain very similar performance concerning accuracy. This result is especially interesting from the perspective of adversarial attacks, as using fewer features for classification can make the system more robust against them (the attacker can manipulate less information to evade the system)~\cite{melis18-eusipco}. The performances attained by System API-based approaches were also better than \texttt{IntelliAV}, which employed a combination of different features (including permissions, user-defined API, and more). System API-based strategies also performed significantly better than \texttt{GreatEatlon}, which based its detection also on information extracted from strings and language properties. Notably, using methods significantly improved the accuracy performances in comparison to packages and classes, in line to what obtained from Experiment $2$.
Finally, we analyzed the performances attained by \texttt{Talos}~\cite{cimitile2017talos}\footnote{We directly obtained Talos from the authors, as it is not currently publicly available.}, a static analysis tool that employs logic rules to perform detection (hence, without machine learning). The attained results were very encouraging, but they strongly depended on the set of rules that have been (manually) established to perform ransomware detection. Conversely, System API-based approaches did not require any manual definition of the detection criteria. Additionally, the analysis times of \texttt{Talos} are significantly slower than the ones attained by methods proposed in this work (an average of $100$ seconds per application, $400$ times slower than ours - see Section~\ref{sec:implementation:subsec:performances}).

\subsection{Experiment 4: Resilience against Obfuscation}
\label{sec:eval:subsec:exp4}

The goal of this experiment was assessing the robustness of System API-based strategies against obfuscated samples, \ie,  understanding whether the application of commercial tools to samples could influence the detection capability of the systems. This evaluation is important, as commercial obfuscation tools are quite popular nowadays since they introduce good protection layers against static analysis (\eg, to avoid pieces of legitimate applications to be copied). Previous works showed that attackers could exploit this aspect by obfuscating malware samples with such tools, thus managing to bypass anti-malware detection~\cite{maiorca15-cose}.  

In this experiment, we primarily focused on obfuscated samples whose original (\ie, non-obfuscated) variant was already included in the training set. Such a choice was made because we wanted to assess if obfuscation was enough to influence the key-features of System API-based methods, thus \emph{changing the classifiers' decision for a sample whose original label was malicious}. 

To this end, we employed a test-bench of ransomware obfuscated with the tool \texttt{DexProtector}\footnote{\url{https://dexprotector.com/}}, a popular, commercial obfuscation suite that allows for protecting Android applications through heavy code obfuscation. Although such a tool is mostly used for legitimate purposes (\eg, protection of intellectual properties), it can also be used by attackers to make malicious applications harder to be detected. Out of the $3017$ ransomware samples, we could obfuscate $2668$ samples (the remaining could not be obfuscated due to errors of the obfuscation software) with three different strategies (for a total of $8004$ obfuscated samples). The strategies employed to obfuscate samples were the following: 

\begin{itemize}
	\item \textbf{String Encryption}. This strategy encrypts strings that are related to \texttt{const-string} instructions, and injects a user-implemented method that performs decryption at runtime. 
	\item \textbf{Resource Encryption}. It encrypts the external resources contained in the \texttt{res} and \texttt{assets} folders. To do so, it adds System API information to the \texttt{classes.dex} file, in order to properly manage the encryption routines. 
	\item \textbf{Class Encryption}. This strategy encrypts user-implemented classes, and injects routines that allow to perform dynamic loading of such classes. 
\end{itemize}

\begin{figure}
	\centering
	\includegraphics[width=\linewidth]{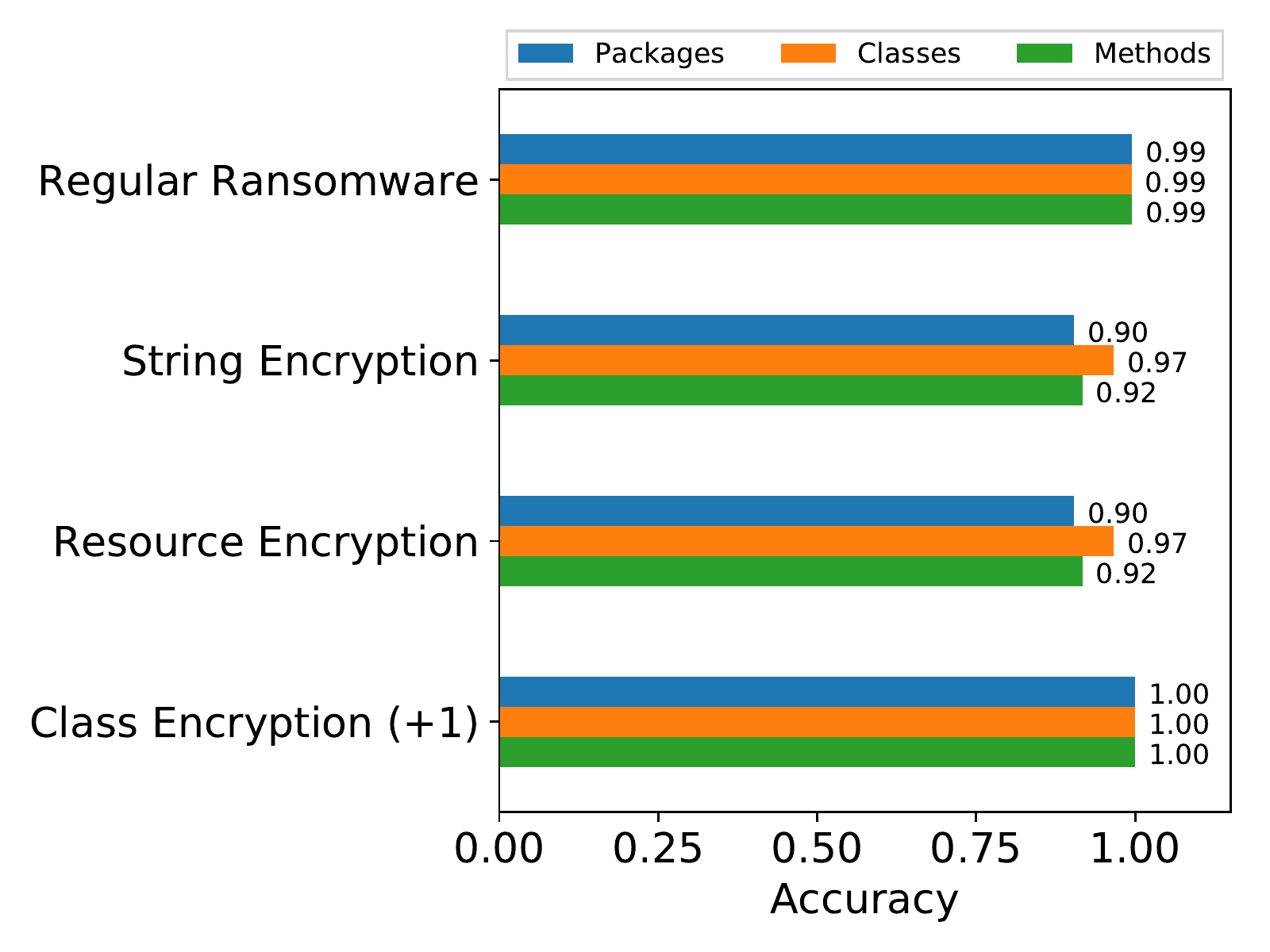}
	\caption{Accuracy performances attained on ransomware samples that have been obfuscated with three different techniques. The accuracy values are reported for the three System API-based detection.}
	\label{sec:eval:subsec:exp4:fig:obfuscation}
\end{figure}

Figure \ref{sec:eval:subsec:exp4:fig:obfuscation} reports the accuracy attained by the three System API-based strategies against the obfuscated samples. Such results show that all the detection strategies (without significant differences between each other) are resilient against obfuscation attempts. However, Class Encryption deserves separate consideration. This strategy employs heavy obfuscation, and it was explicitly performed to defeat static analysis. Typically, none of the static-based techniques that analyze the executable file should be able to detect such attacks correctly. However, this obfuscation strategy introduces a very regular sequence of System API-based routines that manage runtime decryption of the executable contents. 

For this reason, it is sufficient to inject only one sample inside the training set to make all obfuscated samples to be detectable. Hence, we added the $+1$ mark to Class Encryption. Notably, this may create false positives when legitimate samples are obfuscated with the same strategy. Nevertheless, it is sporadic to find such applications, as Class Encryption strongly decreases the application performances \cite{maiorca15-cose}, and much simpler obfuscation techniques are generally used.

\section{Implementation and Performances}
\label{sec:implementation}
Although many solutions have been proposed in the wild to detect ransomware and generic malware, almost none (with the exception, for example, of \cite{ahmadi17-cdmake}) was ported to Android devices, often due to the complexity of the proposed approaches. However, an offline, on-device solution is very useful to perform early detection of applications downloaded, for example, from third-party markets (which are more subjected to malware attacks). For this reason, and also to demonstrate the suitability of System API-based approaches, we ported the simplest of the three strategies (Package-based) with the name of \texttt{R-PackDroid} (as it implements the same approach introduced in our previous work \cite{maiorca17-sac}). This implementation scans for any downloaded, installed and updated applications, and it classifies them as ransomware, malware or legitimate. If an application is found as malicious, the user can immediately remove it. 

\texttt{R-PackDroid} has been designed to work on the largest amount of devices possible. Hence, during its development, we focused on optimizing its speed and battery consumption. For this reason, we avoided any textual parsing of bytecode lines (which can be attained by transforming the \texttt{.dex} file to multiple \texttt{.smali} files with \texttt{ApkTool}). Therefore, we resorted to \texttt{DexLib}, a powerful parsing library part of the \texttt{baksmali}\footnote{\url{https://github.com/JesusFreke/smali}} disassembler (and used by \texttt{ApkTool} itself), to directly extract method calls and their related packages. This library allowed to obtain a very high precision at analyzing method calls and significantly reduces the presence of bugs or wrong textual parsing in the analysis phase.

The classification model has been implemented by using \texttt{Tensorflow}\footnote{\url{https://www.tensorflow.org/}}, an open source, machine learning framework which has been designed to be also used in mobile phones. In particular, we adapted its Random Forest implementation (\texttt{TensorForest}) to the Android operating system. Notably, our Android application only performs classification by using a previously trained classifier. The training phase is carried out separately, on standard X86 architectures. This choice was made to ensure the maximum easiness of use to the final user, thus reducing the risk of invalidating the existing model. 

\begin{figure}[htp]
	\centering
	\includegraphics[width=.75\linewidth]{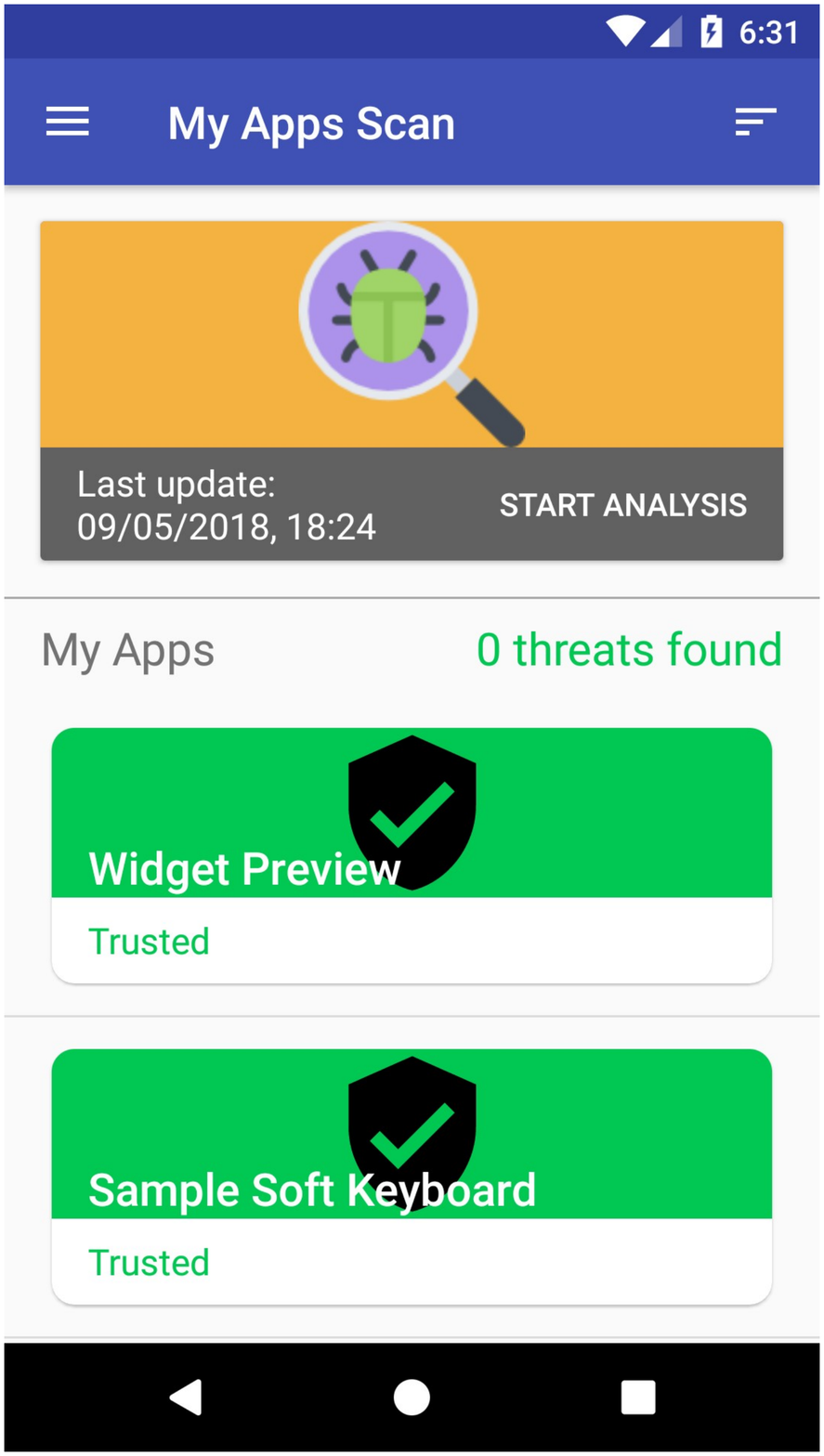}
	\caption{An example of the Android \texttt{R-PackDroid} screen.}
	\label{sec:method:fig:rpack}
\end{figure}

Figure \ref{sec:method:fig:rpack} shows an example of the main screen of \texttt{R-PackDroid}. The application is parsed either when it is downloaded from any store, or when the user decides to scan it (or to scan the whole file system). Each application is identified by a box, whose color is associated with the application label (green for trusted, red for malware and violet for ransomware). After getting the result, by clicking on each box related to the scanned application, it is possible to read more details about the packages that it employs, as well as general information such as the app hash and size. Moreover, if the user believes that the result reported by \texttt{R-PackDroid} is wrong, she can report it by simply tapping a button (a privacy policy to accept is also included). To this scope, we resort to the popular service \texttt{FireBase}\footnote{\url{https://firebase.google.com/}}. \texttt{R-PackDroid} is available for free on the Google Play Store (for the moment, Android versions until $7.1$ are supported). 

\subsection{Computational Performances}
\label{sec:implementation:subsec:performances}
\begin{figure}[htp]
	\centering
	\includegraphics[width=\linewidth]{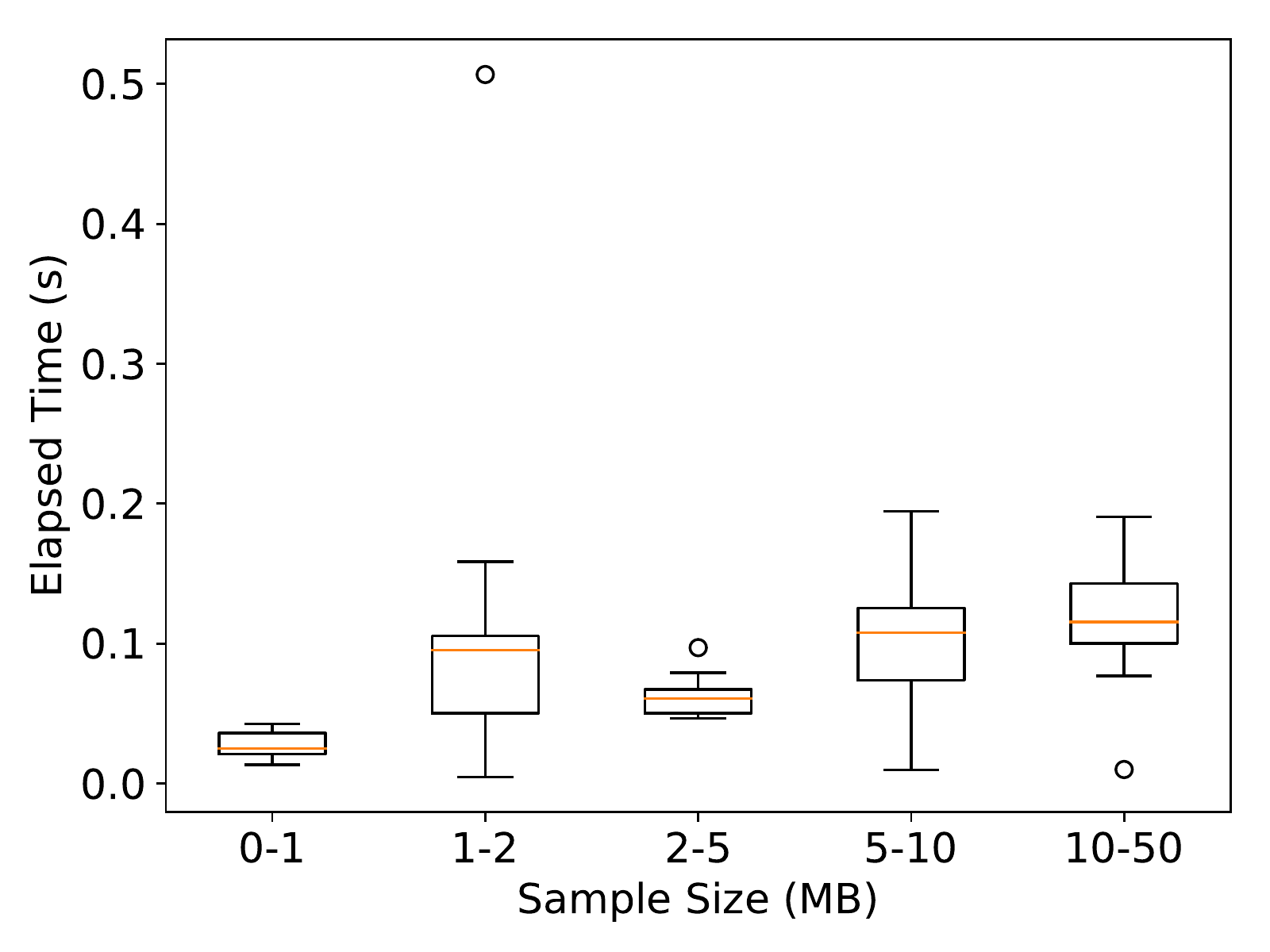}
	\caption{Analysis performances on a X86 workstation, with the elapsed time in seconds, for different \texttt{.apk} sizes.}
	\label{sec:eval:subsec:performances:fig:vortexboxplot}
\end{figure}

We analyzed the computational performances of \texttt{R-PackDroid} by running it both on X86 and Android environments. In particular, we focused on extracting the time interval between the \texttt{.apk} loading and the generation of the feature vector for $100$ benign samples (grouped by their \texttt{.apk} size)\footnote{The elapsed time to classify a sample, \ie, to read its feature vector and get the final label, is negligible.}. The choice of benign samples was made because they are typically more complex to be analyzed in comparison with generic malware and ransomware. We first ran our experiments on a $24$-core Xeon machine with $64$ GB of RAM. The attained results, shown in Figure \ref{sec:eval:subsec:performances:fig:vortexboxplot}, proved that our system could analyze even huge applications in less than $0.2$ seconds.

\begin{figure}[htp]
	\centering
	\includegraphics[width=\linewidth]{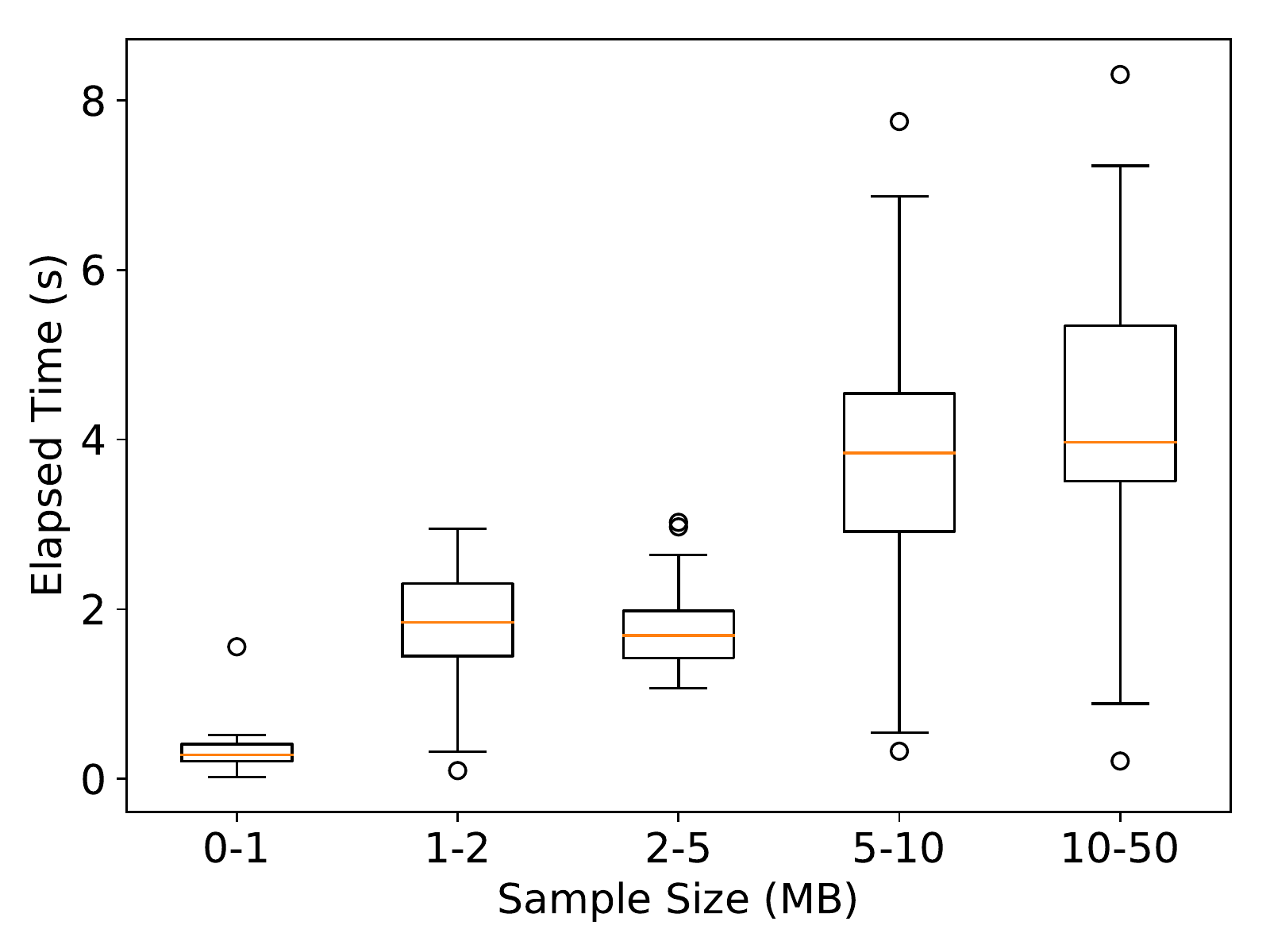}
	\caption{Analysis performances on a real device, with the elapsed time in seconds, for different \texttt{.apk} sizes.}
	\label{sec:eval:subsec:performances:fig:nexusboxplot}
\end{figure}

To evaluate the performances of \texttt{R-PackDroid} on a real Android phone, we ran the same analysis on a Nexus 5, a $5$-years-old, $4$-core device with $2$ GB of RAM, equipped with the 6.0.1 version of Android.  
Results are reported in Figure \ref{sec:eval:subsec:performances:fig:nexusboxplot}. Even if the analysis times were slower than X86 machines, and even if we were using, in this case, the slowest version of the algorithm, the average analysis time for very large apps was slightly more than $4$ seconds. This result was very encouraging, and it showed that \texttt{R-PackDroid} could be safely used even on old phones. The higher dispersion of the time values, in comparison to the ones attained in the previous picture, was possibly caused by the presence of other background processes in the device.

Finally, it is also important to observe that the analysis time is not strictly proportional to the \texttt{.apk} size, as the file may contain additional resources (\eg, images) that increase the \texttt{.apk} size, without influencing the size of the \texttt{DexCode} itself. For this reason, it was not surprising to see the attained average values did not necessarily increase with the \texttt{.apk} size.

\section{Discussion and Limitations}
\label{sec:discussion}

The results attained in Sections \ref{sec:eval} and \ref{sec:implementation} can be summarized with the following findings:
\begin{itemize}
	\item \textbf{Finding 1}. System API-based information could be effectively used, \emph{alone}, to properly distinguish ransomware from generic malware and legitimate applications.
	\item \textbf{Finding 2}. Using finer-grained information (classes and methods), albeit involving more features in the analysis, brought significant improvements to accuracy when detecting previously, unseen samples. Moreover, using API-methods allowed for more accuracy under low false positives values.
	\item \textbf{Finding 3}. System API-based approaches could obtain comparable performances to other approaches that involved more features of different types. 
	\item \textbf{Finding 4}. System API-based approaches guaranteed robustness against typical obfuscation strategies such as string encryption. However, by including a few obfuscated samples in the training set, it was also possible to detect heavy, anti-static obfuscation techniques such as class encryption. 
	\item \textbf{Finding 5}. System API-based approaches were well suitable to be ported and implemented on mobile devices, with excellent computational performances even on very large applications. 
\end{itemize}

We point out that it would be possible to evade the proposed System API-based approaches by replacing System-related packages/classes/methods with semantically equivalent, user-implemented ones. For example, attackers may have two possibilities to replace System API-based methods: \emph{(i)} creating copies of the original instructions of the methods and injecting them into fake methods; \emph{(ii)} re-implementing the methods by using customized instructions/logic. However, these two approaches may feature some critical limitations. In the first approach, the attacker is forced to import the copies of the instructions to the \texttt{dex} code (as the methods become user-implemented). However, the imported codes may contain further references to other System API-based methods, which would need to be replaced. Therefore, this procedure may become unfeasible, considering the high variety of calls that can be invoked. The second approach may be hard to implement if the methods to be replaced are very sophisticated (\eg, methods related to cryptography or the execution of activities). We observe that the first approach can be more effective when replacing System API-based packages, as their variety is significantly lower in comparison to methods.  
   
There would also be the possibility that a skilled attacker attempts to evade System API-based detection algorithms by performing Adversarial Machine Learning attacks, such as test-time evasion~\cite{biggio13-ecml,demontis17-tdsc}. In this scenario, the goal would be evading the classifier detection with a minimal number of changes by performing fine-grained modifications to the features of the analyzed test samples. However, this strategy may be challenging to be performed in practice. The problem, also known in the literature as \emph{Inverse Feature Mapping}~\cite{biggio13-ecml,biggio14-ijprai,biggio18-pr}, is constructing the real sample that implements the modifications made to the feature vectors. As the changes to the feature vector would involve the injection or removal of specific System API-based information, they may not be feasible for the reasons we mentioned in the previous paragraph. We plan to inspect the adversarial-related aspects of System API-based methods, as well as the practical creation of evasive samples, in future work.

It is also worth noting that since Android Oreo ($8.0$), Google introduced new defenses against background processes that are typical of ransomware (\eg, the ones that directly lock the device). However, this does not exclude other malicious actions on the application level. For this reason, it is always better to have an additional system that can detect attempts at performing malicious actions. 

Finally, we also point out that, during our tests, we found samples that could not be analyzed due to crashes and bugs of the \texttt{DexLib} library, and that have therefore been excluded from our analysis. However, their percentage (regarding the whole corpus that we analyzed) is negligible (less than $1\%$ of the whole file corpus).

\section{Conclusions}
\label{sec:conclusions}

In this work, we provided a detailed insight into how System API-based information could be effectively used (also on a real device) to detect ransomware and to distinguish it from legitimate samples and generic malware. The attained experimental results demonstrated that, by using a compact set of information tailored to the detection of a specific malware family (ransomware), it was possible to achieve detection performances (also on other malware families) that were comparable to systems that employed a much more complex variety of information. Moreover, System API-based information also proved to be valuable to detect obfuscated samples that focused on hiding user-implemented information. Notably, although it is tempting to combine as many information types as possible to detect attacks (or to develop computationally heavy approaches), it may not be the only, feasible way to construct accurate, reliable malware detectors. For this reason, we claim that future work should focus on developing reliable, small sets of highly discriminant features that cannot easily be manipulated by attackers (with a particular reference to machine learning attacks). Moreover, a clear understanding of the impact of each feature on the classifier detection (also known as \emph{explainability}) can help analysts understand the classifiers errors and to improve their detection capabilities.

\section*{Acknowledgements}
This work was partially supported by the following projects: {\it NeCS} (H2020 EU funded - GA \#675320); {\it C3ISP} (H2020 EU funded - GA \#700294); {\it INCLOSEC} (funded by Sardegna Ricerche - CUPs G88C17000080006); {\it PISDAS} (funded by Regione Autonoma della Sardegna - CUP E27H14003150007). The authors also thank Marco Lecis for his valuable contribution to the paper experiments. 

\bibliographystyle{model1-num-names}
\bibliography{android,misc}

\end{document}